\documentclass[draft]{agujournal2019}
\usepackage{url} 
\usepackage{lineno}
\usepackage[inline]{trackchanges} 
\usepackage{soul}

\draftfalse

\journalname{Geophysical Research Letters}

\usepackage{pdflscape,epstopdf,textgreek,tabularx,colortbl}
\usepackage[shortlabels]{enumitem}
\newcolumntype{L}{>{\raggedright\arraybackslash}X}
\usepackage{lmodern}
\usepackage[T1]{fontenc}
\usepackage{textcomp}

\begin{document}


\title{Detection of a bolide in Jupiter's atmosphere with Juno UVS}


\authors{Rohini S. Giles\affil{1}, Thomas K. Greathouse\affil{1}, Joshua A. Kammer\affil{1}, G. Randall Gladstone\affil{1,2}, Bertrand Bonfond\affil{3}, Vincent Hue\affil{1}, Denis C. Grodent\affil{3}, Jean-Claude G\'{e}rard\affil{3}, Maarten H. Versteeg\affil{1}, Scott J. Bolton\affil{1}, John E. P. Connerney\affil{4,5} and Steven M. Levin\affil{6}}

\affiliation{1}{Space Science and Engineering Division, Southwest Research Institute, San Antonio, Texas, USA}
\affiliation{2}{Department of Physics and Astronomy, University of Texas at San Antonio, San Antonio, Texas, USA}
\affiliation{3}{Laboratoire de Physique Atmosphérique et Planétaire, STAR Institute, Université de Liège, Liège, Belgium}
\affiliation{4}{Space Research Corporation, Annapolis, Maryland, USA}
\affiliation{5}{Goddard Space Flight Center, Greenbelt, Maryland, USA}
\affiliation{6}{Jet Propulsion Laboratory, Pasadena, California, USA}

\correspondingauthor{R. S. Giles}{rgiles@swri.edu}

\begin{keypoints}
\item Juno UVS recorded transient blackbody emission from a point source in Jupiter's atmosphere
\item The emission is consistent with a fireball produced by a 250--5000 kg impactor in Jupiter's upper atmosphere
\item We estimate an impact flux on Jupiter of 24,000 per year for masses greater than 250--5000 kg
\end{keypoints}


\begin{abstract}

The UVS instrument on the Juno mission recorded transient bright emission from a point source in Jupiter's atmosphere. The spectrum shows that the emission is consistent with a 9600-K blackbody located 225 km above the 1-bar level and the duration of the emission was between 17 ms and 150 s. These characteristics are consistent with a bolide in Jupiter's atmosphere. Based on the energy emitted, we estimate that the impactor had a mass of 250--5000 kg, which corresponds to a diameter of 1--4 m. By considering all observations made with Juno UVS over the first 27 perijoves of the mission, we estimate an impact flux rate of 24,000 per year for impactors with masses greater than 250--5000 kg. 
    
\end{abstract}

\section*{Plain Language Summary}

UVS is an ultraviolet spectrograph on NASA's Juno spacecraft, which has been in orbit around Jupiter since 2016. In April 2020, UVS observed short-lived bright emission from a point source in Jupiter's atmosphere. The emission was consistent with a blackbody of temperature 9600 K. We suggest that this was a fireball produced by a 250--5000 kg meteoroid entering Jupiter's atmosphere.


\section{Introduction}

As the largest and most massive planet in the Solar System, Jupiter undergoes a heavy bombardment of objects, ranging from tiny dust grains to kilometer-sized comets. The largest impacts, such as comet Shoemaker-Levy 9 \cite{asphaug96} and the 2009 impactor \cite{sanchez-lavega10}, occur rarely but leave scars on the planet that can persist for several months \cite{sanchez-lavega98,sanchez-lavega11} and can affect the distribution of trace species in Jupiter's upper atmosphere for decades~\cite{lellouch06,cavalie13,benmahi20}. Impacts from objects in the 5--20 m diameter range occur more frequently and produce short bright flashes of light that can be observed by amateur astronomers on Earth, but no observable debris \cite{hueso18}. Over a period of 8 years (2010--2017), amateur astronomers observed 5 of these smaller impacts, leading to an estimated impact rate of 10--65 per year \cite{hueso18}, an estimate that was not significantly modified by the observation of a 6th impact in 2019 \cite{sankar20}. On the smallest scales, the constant influx of dust particles equates to hundreds of thousands of tons of material per year \cite{sremcevic05,poppe16}.

While larger (\textgreater 5 m) impacts can be observed from Earth, observations from orbiting spacecraft have the advantage of being able to detect the fainter flashes that are produced from the more frequent smaller impacts. One example of this is the small fireball observed by the camera on the Voyager 1 spacecraft, which \citeA{cook81} estimated was caused by an 11 kg (\textless 0.5 m) meteoroid. In this paper, we present observations of another fireball in Jupiter's atmosphere, this time observed by the UVS instrument on the Juno spacecraft. In Section~\ref{sec:observations}, we describe the instrument and the method of observation. In Section~\ref{sec:results} we discuss the properties of the bright blackbody-emission flash observed. In Section~\ref{sec:discussion}, we conclude that this flash was caused by a meteoroid of mass 250--5000 kg (diameter 1--4 m) entering Jupiter's atmosphere, and we use the sum of all of our observations to estimate an impact flux rate. 


\section{Observations}
\label{sec:observations}

\subsection{Juno UVS}

The Ultraviolet Spectrograph (UVS) is an instrument on the Juno spacecraft, which has been in orbit around Jupiter since July 2016 \cite{bolton17}. UVS is a photon-counting imaging spectrograph, covering wavelengths of 68--210 nm in the far-ultraviolet \cite{gladstone17}. The main purpose of UVS is to study the morphology, brightness, and spectral characteristics of Jupiter's auroras, and the instrument's spectral range covers important H and H\textsubscript{2} auroral emissions~\cite{gladstone17b,bonfond17}.

Juno is a spin-stabilized spacecraft with a rotation period of 30 s; as the spacecraft rotates, the UVS instrument slit sweeps across Jupiter. The wavelength, slit position and precise time of UV photon detection events are recorded and this information is then used to build up spatial maps of the ultraviolet radiation (see Figure~\ref{fig:image} for an example from a single spin). The UVS instrument slit consists of two wide segments on either side of a narrow segment. The wide parts of the slit have a width of 0.2$^{\circ}$ and a spectral resolution of 2.0--3.0 nm, while the narrow part of the slit has a width of $\sim$0.025$^{\circ}$ and a spectral resolution of $\sim$1.3 nm \cite{greathouse13}. The spatial resolution along the slit is 0.1--0.3$^{\circ}$. The observations described in this paper were made at the top edge of the upper wide slit, where the spatial resolution is $\sim$0.3$^{\circ}$.

\subsection{Bright flash}

On occasion, the Juno UVS instrument has observed short-lived, localized ultraviolet emission outside of the auroral zone. In \citeA{giles20c}, we described a set of eleven bright transient flashes that were observed with Juno UVS and shared similar characteristics; they lasted $\sim$1.4 ms, they were dominated by H\textsubscript{2} emission and the source region was located $\sim$260 km above the 1-bar level. Based on these characteristics, we concluded that these were likely to be elves, sprites or sprite halos, forms of Transient Luminous Events (TLEs) that occur in the upper atmosphere in response to tropospheric lightning. 

\begin{figure}
    \centering
    \includegraphics[width=12cm]{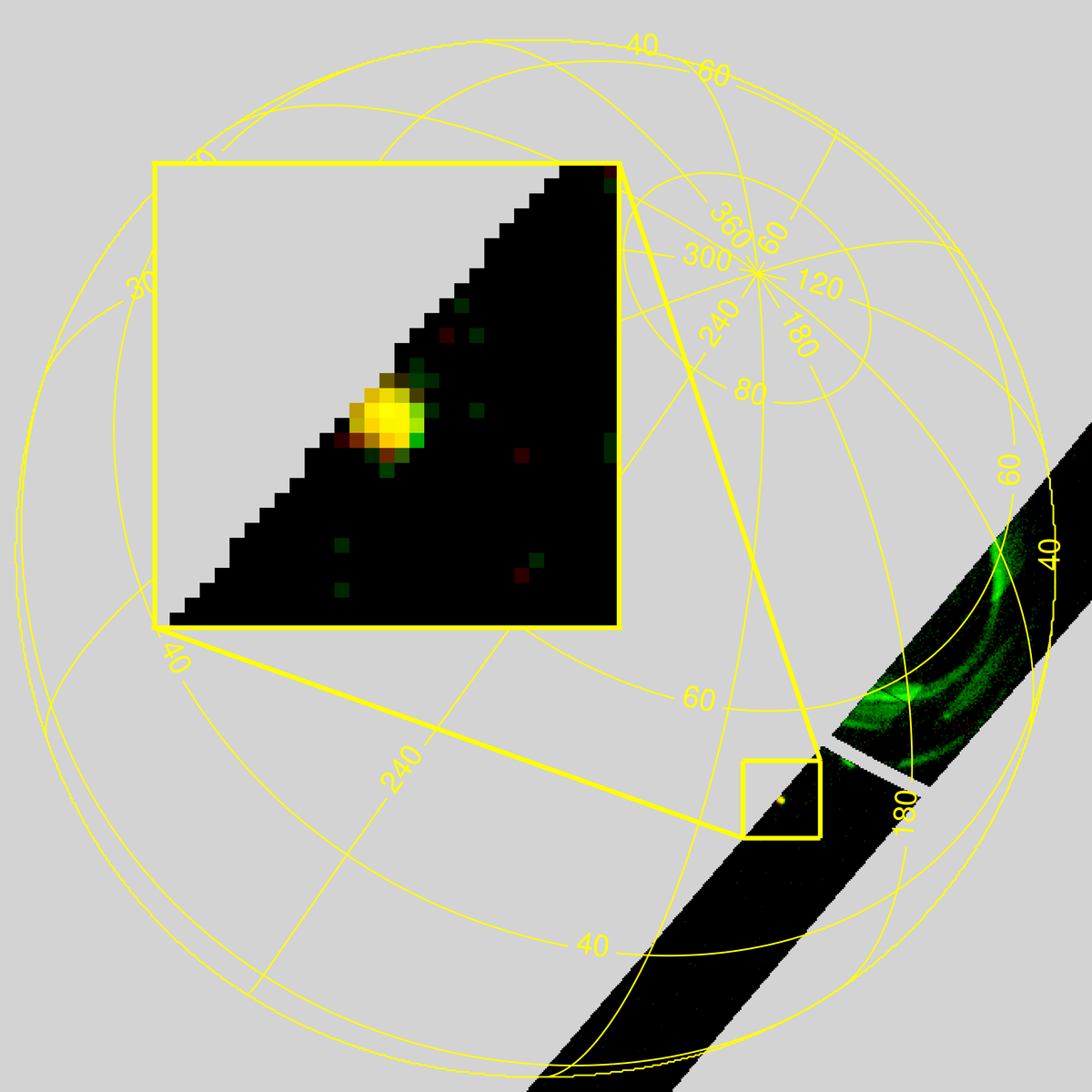}
    \caption{A spatial map of Juno UVS measurements during a single spacecraft spin on 10 April 2020. Photons with wavelengths 130--160 nm are shown in green and photons with wavelengths 170--200 nm are shown in red. The inset image magnifies the area around the bright spot.}
    \label{fig:image}
\end{figure}

In this paper, we focus on another transient bright flash observed by UVS, but this observation has very different characteristics than the aforementioned TLEs. The observation was made on 10 April 2020 at 12:57:10 UTC and the bright spot was located at a planetocentric latitude of 53$^{\circ}$N and a System III longitude of 200$^{\circ}$W. It was observed at an emission angle of 43$^{\circ}$ and a solar incidence angle of 125$^{\circ}$ (i.e. on the planet's nightside). At the time of the observation, the spacecraft was at an altitude of 80,000 km above the 1-bar level of the planet and the sub-spacecraft latitude and longitude were 63$^{\circ}$N and 243$^{\circ}$W respectively. This location was not observed by Juno's infrared instrument \cite<JIRAM,>{adriani08} or the visible light camera \cite<JunoCam,>{hansen17} during this time period or afterwards. Figure~\ref{fig:image} shows the UVS map from the spin in which the bright spot was recorded and includes a magnified image of the spot itself. This magnified image has dimensions 3$^{\circ}\times$3$^{\circ}$ on the sky, with each pixel being 0.1$^{\circ}\times$0.1$^{\circ}$. The spot was only observed during a single spin; it was not seen two spins earlier or three spins later, which were the closest times when the same latitude and longitude was observed.

The colors used in Figure~\ref{fig:image} show the number of photon detections made in two different spectral regions. Green is used to represent photons with wavelengths 130--160 nm and red is used to represent photons with wavelengths 170--200 nm. The UVS swath includes a segment of Jupiter's northern auroral oval and the auroral emission appears purely green in the image; this is because the main auroral H\textsubscript{2} emission bands are at \textless 170 nm. In contrast, the bright spot appears mostly yellow, indicating that there is significant emission at longer wavelengths as well.

Although this bright spot is located close to the northern auroral region, the long-wavelength emission marks it out as unique when compared with auroral features. It is also unique compared to the transient bright flashes described in \citeA{giles20c}, which were all dominated by H\textsubscript{2} emission, like the aurora. Despite these different spectral characteristics, the bright spot does clearly originate from within Jupiter's atmosphere; Figure~\ref{fig:image} shows that it is located far from the limb of the planet so there is no possibility of confusion with a star close to Jupiter, and the bright spot spectrum shows CH\textsubscript{4} absorption from Jupiter's atmosphere (see Section~\ref{sec:spectral}) so it cannot originate from an object located between the spacecraft and the planet. 

We conducted a search for other such events, by filtering for occasions when there was a sharp increase and then decrease in the number of long-wavelength photon counts recorded, but no similar events were found. The spot is analyzed further in Section~\ref{sec:results} and the results are discussed in Section~\ref{sec:discussion}. 


\section{Results}
\label{sec:results}

\subsection{Spatial extent and duration}

Because of the way in which the UVS maps are built up as the spacecraft spins, Figure~\ref{fig:image} contains information about both the spatial extent and the duration of the bright spot. By fitting a two-dimensional Gaussian to the total photon counts, we find that the spot has a FWHM of 0.29$^{\circ}$ in the along slit direction and 0.23$^{\circ}$ in the across slit direction. This is consistent with the shape that we would expect for a point source that has a constant brightness on the 17-ms timescale it takes the slit to pass over the source~\cite{greathouse13}. Based on the distance of the spacecraft at the time of the observations (80,000 km), the upper limit for the spot diameter is 400 km.

Because the shape of the bright spot is consistent with the FWHM of a point source, the bright spot must also be approximately stationary in latitude and longitude over the 17-ms duration of the observation. In order to observe motion, the spot would have had to move \textgreater0.1$^{\circ}$ in 17 ms, equating to a horizontal velocity of \textgreater8000 kms\textsuperscript{-1}. Since we do not observe any elongation, any horizontal movement must be below this speed.

While the brightness appears approximately constant on the timescale it takes the slit to pass over the source, the bright spot was only observed during a single spin, so it must be transient on slightly longer timescales. UVS data was recorded from the same latitude and longitude two spins (60 s) earlier and three spins (90 s) later, and the bright spot was not present in either of these maps. This places an upper bound of 150 s on the duration of the bright spot, giving a duration of between 17 ms and 150 s.

Because the bright spot was observed on the night side of the planet, it is not possible to use UVS data to search for any local atmospheric changes in its aftermath, such as changes in the aerosol abundance. Any such changes could therefore have lasted longer than 150 s.

\subsection{Spectral analysis}
\label{sec:spectral}

\begin{figure}
    \centering
    \includegraphics[width=12cm]{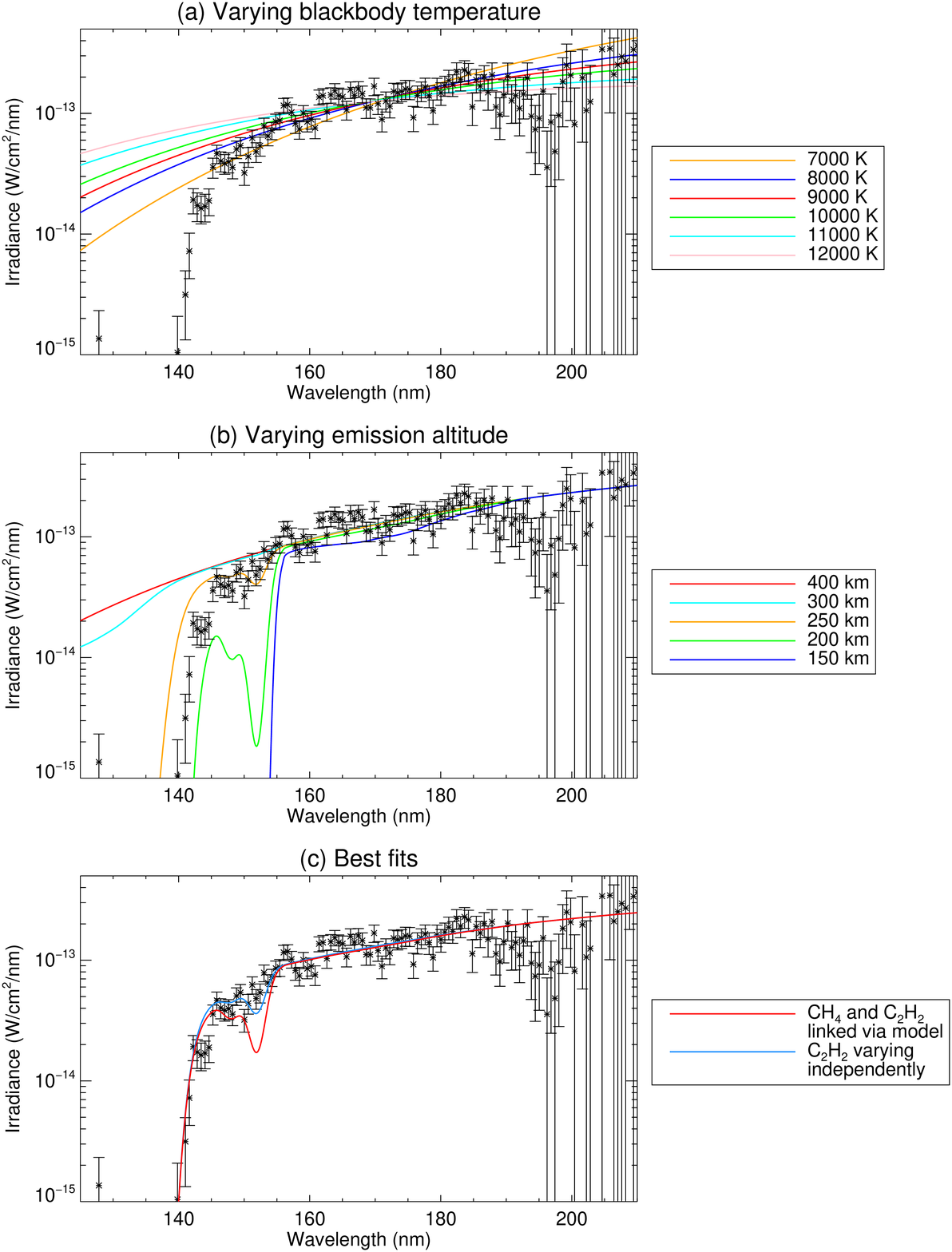}
    \caption{The bright spot spectrum (black) compared with various model spectra. (a) The colored lines show the shape of blackbodies with different temperatures. There is no atmospheric absorption. (b) The colored lines show the spectral shape of a 9000 K blackbody placed at different altitudes and therefore with different amounts of atmospheric absorption. (c) The colored lines show two different best-fit model spectra.}
    \label{fig:meteor_fits}
\end{figure}

The bright spot spectrum was calculated by summing all photon detections within a 1$^{\circ}\times$1$^{\circ}$ box centered on the bright spot. The spectrum is shown by the black data points in Figure~\ref{fig:meteor_fits} and is presented in terms of spectral irradiance measured at the spacecraft. As suggested earlier by the color of the bright spot in Figure~\ref{fig:image}, the bright spot emission has an unusual spectral shape compared to both the auroral emission and the TLE emission described in \citeA{giles20c}. The auroral and TLE spectra are dominated by H\textsubscript{2} emission, and the H\textsubscript{2} Lyman band can be recognized by a double peak at 160 nm. Beyond 170 nm, the H\textsubscript{2} emission drops to low levels. In contrast, the spectrum shown in black in Figure~\ref{fig:meteor_fits} increases smoothly between 150 and 190 nm, suggesting blackbody emission. The rapid drop-off at wavelengths shorter than 140 nm suggests strong atmospheric absorption by CH\textsubscript{4}. 

In order to model a blackbody spectrum attenuated by atmospheric absorption, we used the atmospheric composition Model C from~\citeA{moses05}. In this model atmosphere, the CH\textsubscript{4} homopause is located at a pressure of $2.9\times10^{-4}$ mbar, or 360 km above the 1 bar level. For a given altitude and emission angle, we used the atmospheric composition model to calculate the column density for each atmospheric gas. The three stratospheric gases that contribute significantly to absorption in the 125--210 nm range are CH\textsubscript{4}, C\textsubscript{2}H\textsubscript{2} and C\textsubscript{2}H\textsubscript{6}. Absorption cross-sections for these three gases were obtained from~\citeA{chen04},~\citeA{lee01},~\citeA{smith91} and~\citeA{benilan00}. The atmospheric transmission was then calculated using the column density for each gas and the absorption cross-sections, and this was multiplied by a blackbody spectrum to obtain a modeled top-of-atmosphere spectrum. The colored lines in Figures~\ref{fig:meteor_fits}(a) and~\ref{fig:meteor_fits}(b) show these modeled spectra for a range of blackbody temperatures and emission altitudes. 

Figure~\ref{fig:meteor_fits}(a) shows modeled spectra for blackbodies with five different temperatures. For these spectra, no atmospheric absorption is included and the spectra have been scaled to match the average irradiance of the bright spot spectrum. Within the 125--210 nm spectral range, increasing the temperature from 7000 K to 12,000 K flattens the spectrum. At shorter wavelengths, the data diverges sharply from all of the blackbody spectra due to the lack of atmospheric absorption. We note that the data also diverges from the blackbody spectra at $\sim$195 nm. As demonstrated by the larger error bars, the UVS sensitivity is much lower at these longer wavelengths than at shorter wavelengths; by 195 nm, the instrument's effective area is an order of magnitude smaller than at 160 nm \cite{hue18}. We therefore attribute this apparent small dip in the irradiance to limitations in the radiometric calibration. 

Figure~\ref{fig:meteor_fits}(b) shows modeled spectra obtained using different emission altitudes (i.e. different amounts of atmospheric absorption). Each spectrum uses the same 9000 K blackbody as the source emission and the altitude is defined as the height above the 1-bar level. The red lines in Figures~\ref{fig:meteor_fits}(a) and ~\ref{fig:meteor_fits}(b) are identical, as there is negligible absorption above an altitude of 400 km. As the source is moved deeper in the atmosphere, the amount of absorption increases. By 250 km there is significant CH\textsubscript{4} absorption shortwards of 140 nm. By 200 km, there is also a significant C\textsubscript{2}H\textsubscript{2} feature at 152 nm. By 150 km, the atmosphere is opaque shortwards of 155 nm due to the combination of C\textsubscript{2}H\textsubscript{2}, C\textsubscript{2}H\textsubscript{6} and CH\textsubscript{4}.

When the blackbody temperature, emission altitude and a scaling factor were all allowed to vary together, the best fit was obtained with a temperature of 9600$\pm$600 K and an altitude of 225$\pm$5 km above the 1-bar level (assuming an error of 20\% on the abundances of the absorbers in the model). This best-fit spectrum is shown by the red line in Figure~\ref{fig:meteor_fits}(c). While this modeled spectrum provides a good fit at most wavelengths, the C\textsubscript{2}H\textsubscript{2} absorption feature at 152 nm is clearly too strong; an altitude of 225 km provides a good fit for the CH\textsubscript{4} and C\textsubscript{2}H\textsubscript{6} column densities, but the  C\textsubscript{2}H\textsubscript{2} column density ($2.1\times10^{16}$ cm\textsuperscript{-2}) is too high. Instead, if we allow the C\textsubscript{2}H\textsubscript{2} column density to vary independently of the other gases, we retrieve a value of $1.1\times10^{16}$ cm\textsuperscript{-2}, approximately half of the value at 225 km. The spectrum obtained with this value is shown by the blue line in Figure~\ref{fig:meteor_fits}(c). Compared to the~\citeA{moses05} model atmosphere, this suggests that the atmosphere near the bright spot has lower C\textsubscript{2}H\textsubscript{2} volume mixing ratio. This is unsurprising as the~\citeA{moses05} model describes equatorial latitudes, and~\citeA{nixon07} used Cassini CIRS observations to show that the C\textsubscript{2}H\textsubscript{2} volume mixing ratio decreases at higher latitudes. The blue line in Figure~\ref{fig:meteor_fits}(c) is therefore consistent with an altitude of 225 km at a latitude of 53$^{\circ}$N. 

\section{Discussion}
\label{sec:discussion}

The bright spot described in Section~\ref{sec:results} has the following properties:

\begin{enumerate}[(i)]

    \item Its duration is longer than 17 ms and less than 150 s.
    
    \item The shape is consistent with a point source, which equates to a maximum diameter of 400 km. 
    
    \item Its spectral shape is consistent with a blackbody of temperature 9600 K located at an altitude of 225 km above the 1-bar level (a pressure of 0.04 mbar).
    
\end{enumerate}

\begin{figure}
    \centering
    \includegraphics[width=12cm]{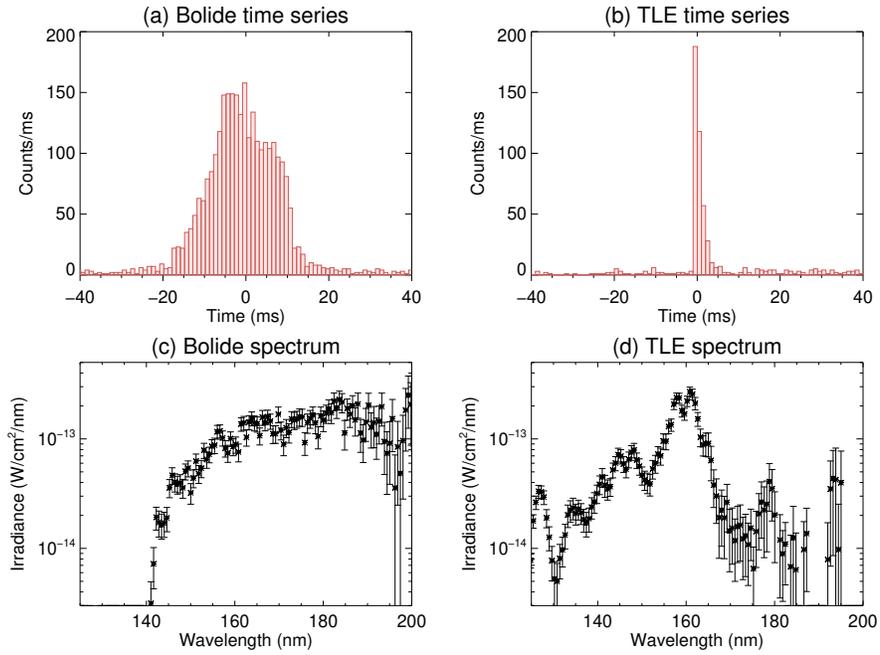}
    \caption{A comparison of the time series and spectra of the bolide (this paper) and the TLEs \cite{giles20c}. (a) and (b) show the number of observed UV photons as a function of time, relative to the peak event time. The time series shown in (b) is for the TLE observed on April 10 2020 at 17:24:35 UTC. (c) shows the same spectrum presented in Figure~\ref{fig:meteor_fits} and (d) shows the averaged TLE spectrum from \citeA{giles20c}.}
    \label{fig:meteor_tle}
\end{figure}

In \citeA{giles20c}, we presented several instances of transient bright flashes observed in the Juno UVS data and we concluded that these were consistent with TLEs known as elves, sprites or sprite halos, upper atmospheric responses to tropospheric lightning. The bright spot that we present in this paper differs from the previous bright flashes in two main ways, highlighted in Figure~\ref{fig:meteor_tle}. Firstly, it has a significantly longer duration. The TLEs had a duration of $\sim$1.4 ms, which is short enough that UVS could observe their decay during the time it took the instrument slit to pass across the point source. In contrast, this bright spot is approximately constant as the slit passes across it, giving it a minimum duration of $\sim$17 ms. This difference in duration is shown in Figures~\ref{fig:meteor_tle}(a) and (b); the Gaussian shape in (a) is governed by the width of the slit, while the rapid exponential decay of (b) takes place on a much shorter timescale. Secondly, the spectra of the TLEs were dominated by H\textsubscript{2} Lyman band emission, giving them a very similar spectral shape to auroral emission. In contrast, this bright spot has a blackbody-like emission spectrum. The spectra are compared in Figures~\ref{fig:meteor_tle}(c) and (d). As discussed in Section~\ref{sec:results}, the shape of (c) is consistent with blackbody emission, while the double peak at 160 nm seen in (d) is indicative of H\textsubscript{2} emission. 

Because of both the duration and the spectral shape, the bright spot described in this paper is unlikely to be a TLE. Lightning and associated TLEs do not have blackbody spectra; on earth, they have emission spectra that are dominated by nitrogen and oxygen lines~\cite{walker17,rodger99} and on Jupiter their emission spectra are dominated by hydrogen~\cite{borucki96,yair09}. Elves have sub-millisecond timescales, and while sprites can last for several tens of milliseconds, the brighter ones are often shorter~\cite{lyons06}. There are other longer-lasting TLEs, such as blue jets, but these would still be H\textsubscript{2} dominated and they emerge directly from the thunderstorm anvils~\cite{rodger99}, so we would expect them to be deeper in the atmosphere. The shape of the spectrum also rules out the possibility that this is a transient auroral event, as auroral emission is H\textsubscript{2} dominated.

Given the spectrum and the duration, we suggest that the bright spot we observe may be a bolide/fireball in Jupiter's atmosphere. \citeA{borovicka96} studied the lightcurves and spectra of two of the brightest bolides observed in the Earth's atmosphere and found that the spectra were a mixture of blackbody continuum and emission lines. The emission lines come from ablated meteoroid material and the blackbody continuum comes from thermal bremsstrahlung (free-free) emission. In the beginning phase of the bolide, \citeA{borovicka96} found that the line emission dominated, but during the brightest part of the lightcurve, the blackbody continuum dominated. A temperature of 9600 K is approximately consistent with the Planck temperatures measured in Earth bolides. \citeA{borovicka94} found that fireball spectra in the Earth's atmosphere exhibit two distinct characteristic temperatures: a primary component with a brightness temperature of 4000 K and a secondary component with a brightness temperature of 10,000 K. The lower temperature is thought to be from the thermal radiation of the meteoroid itself, while the second part is associated with the shock wave. The greater the speed of the meteoroid, the more dominant this second part is. We do not observe any movement of the bright spot during our 17-ms observation timescale, and this is consistent with a typical impact velocity of tens of kms\textsuperscript{-1}~\cite{crawford94}; a horizontal velocity of \textgreater8000 kms\textsuperscript{-1} would be required to be detectable by UVS.

Our observations are somewhat similar to the observations made by the UVS instrument on the Galileo spacecraft when comet Shoemaker-Levy 9 collided with Jupiter in 1994~\cite{hord95}. They observed a brightening during one spatial scan only (no brightening in scans obtained 5 s beforehand or 5 s afterwards) and when combined with simultaneous observations from the Photopolarimeter Radiometer, they concluded that the brightness temperature was $7800^{+500}_{-600}$ K. Our observations are also similar to several observations made by amateur observers of bolides on Jupiter~\cite{hueso18}. These observations were recorded in the visible and the observed bright flashes lasted 1--2 s in each case. One of these events was simultaneously recorded in two different filters, and fitting a blackbody to these two measurements gave a temperature of 6500--8500 K \cite{hueso13}. 

\citeA{hueso10} used the Earth-based observation of a Jovian bolide to estimate the size of the impacting object and we follow their analysis approach here. If we correct for atmospheric absorption, extend the blackbody shape to all wavelengths and integrate, the total irradiance observed at the spacecraft would be $1.2-2.3\times10^{-10}$ Wcm\textsuperscript{-2} (taking into account the 600 K error on the brightness temperature). Assuming isotropic radiation, the total power emitted is $1.0-1.9\times10^{11}$ W. As we only observe the bright flash for a very short period of time as the spacecraft spins, we do not know the total duration of the emission and this adds significant uncertainty to the calculation. Here, we assume a duration of 1--2 s, based on \citeA{hueso18}. There is additional uncertainty from the fact that we do not know which stage of the light curve we are observing - it could be the peak of the emission, or the peak could occur shortly before/after our observations. Including a 50\% uncertainty for this factor, we obtain a total optical energy of $5\times10^{10}$ -- $6\times10^{11}$ J. 

\citeA{brown02} found that the efficiency, $\mu$, with which the kinetic energy of an impactor is converted into optical energy can be empirically described by

\begin{equation}
    \mu = 0.121\times E_{0}^{0.115}
\end{equation}

where $E_{0}$ is the optical energy in terms of kilotons of TNT (1 kton = $4.185\times10^{12}$ J). For our observed total optical energy range, the efficiency is 7--10\% and the kinetic energy of the impactor is therefore $7\times10^{11}$-- $6\times10^{12}$ J. Assuming an impactor velocity of 60 kms\textsuperscript{-1}, the velocity at which fragments of Shoemaker-Levy 9 collided with Jupiter~\cite{crawford94}, and a velocity error of 20\% leads to a mass estimate of 250--5000 kg. For densities of of 250--2000 kgm\textsuperscript{-3}, this equates to a diameter of 1--4 m.

By considering the total power emitted and the blackbody temperature, we can use the Stefan-Boltzmann law to calculate the area of the emitting region. Assuming an emissivity of 1, the effective diameter of the emitting region is 9 m. This is consistent with an impactor diameter of 1--4 m; \citeA{borovicka96} found that the maximum diameter of the radiating region is on the order of 10 times larger than the initial diameter of the body.

A mass estimate of 250--5000 kg seems broadly consistent with the altitude at which we observe the bright flash: 225 km above the 1-bar level, or a pressure of 0.04 mbar. Numerical impact simulations have been used to estimate the altitudes of bolides in Jupiter's atmosphere, and have found that impactors with masses of $\sim1\times10^{6}$ kg reach their peak brightness at 60--150 km above the 1 bar level \cite{hueso13,sankar20}. Smaller impactors, such as our observation, would not penetrate as deeply into the atmosphere. \citeA{borovicka96} studied the entry of a 5000-kg impactor in the Earth's atmosphere, equivalent to the upper end of our mass estimate. They found that the peak brightness occurred at an altitude of 67 km, which corresponds to a pressure of 0.07 mbar \cite{standardatmosphere}. This is slightly deeper in the atmosphere than our observed bright flash pressure level of 0.04 mbar, which is consistent with 5000 kg being the upper bound of our mass estimate. 

By considering all observations made with Juno UVS over the first 27 perijoves of the mission, we find that UVS obtained a total effective coverage of $8.2\times10^{13}$ km\textsuperscript{2}s. In that time, there was a single bolide observation. There are clear limitations in calculating an impact flux rate from a single observation, but a maximum likelihood estimation calculation leads to an impact rate of $\sim$24,000 per year (See Text S1 in the Supporting Information). By considering how impact rates vary as a function of meteoroid size, we can compare this impact rate with the rates deduced from previous studies. We find that our rate is lower than the flux rate estimated by \citeA{cook81} and higher than the flux rate of \citeA{hueso18} (See Text S2 in the Supporting Information). This comparison is dependent on the assumed mass of our observed meteor, and the lower limit of 250 kg is significantly more consistent with \citeA{hueso18} than the upper limit of 5000 kg. We note that there are considerable uncertainties in our estimated rate, as we have thus far only observed a single bolide with Juno UVS.

\section{Conclusions}

Juno UVS observations recorded transient blackbody emission from a point source in Jupiter's atmosphere. Spectral modelling showed that the emission is consistent with a 9600 K source located 225 km above the 1-bar level, and the emission lasted between 17 ms and 150 s. The blackbody nature of the spectrum, the temperature and the duration of the emission are all consistent with a bolide in Jupiter's atmosphere. Based on the energy emitted, we estimate that the impactor had a mass of 250--5000 kg, which would correspond to a diameter of 1--4 m. This impactor size is larger than the small fireball observed by \citeA{cook81} and smaller than the superbolides described by \cite{hueso18}. We estimate an impact flux rate of 24,000 per year, for masses of 250--5000 kg or greater; if we use the lower limit of our mass estimate, this is consistent with the rate determined by \citeA{hueso18}. However, we note that there are large uncertainties associated with this value due to the fact that we have thus far only observed one bolide with Juno UVS. As the Juno mission continues, our integration time on the planet will continue to steadily increase and our estimate of the impact flux rate will improve.

\section*{Acknowledgements}

We are grateful to NASA and contributing institutions, which have made the Juno mission possible. This work was funded by NASA's New Frontiers Program for Juno via contract with the Southwest Research Institute. B.B. is a Research Associate of the Fonds de la Recherche Scientifique - FNRS.

\section*{Data Availability Statement}

The Juno UVS data used in this paper are archived in NASA's Planetary Data System Atmospheres Node: https://pds-atmospheres.nmsu.edu/PDS/data/jnouvs\_3001 \cite{trantham14}. The data used to produce the figures in this paper are available in \citeA{giles21}.

\nocite{bland06,eliason93}

\clearpage

\Large
\noindent\textbf{Supporting Information for ``Detection of a bolide in
Jupiter's atmosphere with Juno UVS''}
\normalsize

\vspace{1cm}

\noindent\textbf{Contents of this file}
\begin{enumerate}
\item Text S1: Impact rate
\item Text S2: Comparison of impact rate with previous studies
\end{enumerate}

\noindent\textbf{Introduction}

This supporting information contains details about the impact flux rate calculation and the comparison to previous impact flux rate estimates.

\noindent\textbf{Text S1: Impact rate}

By multiplying the time spent observing Jupiter by the footprint of the UVS slit on the planet, we find that over 27 perijoves, UVS obtained coverage of $9.3\times10^{11}$ km\textsuperscript{2}s. However, a given point on the planet is only observed for 17 ms as the instrument slit passes over it, and the bright flash caused by a meteor typically lasts 1--2 s. Therefore, the effective coverage considered in the search for bright flashes must be 1.5/0.017 = 88 times larger, giving a total effective coverage of $8.2\times10^{13}$ km\textsuperscript{2}s. Using the surface area of the planet, this is the equivalent of observing the entirety of Jupiter for 1300 seconds (although we note that our observations are not evenly distributed across the planet).

During that time period, we observed one bolide. It is not possible to calculate an accurate estimate of the average occurrence rate from a single event. However, we can use maximum likelihood estimation \cite{eliason93} to find the occurrence rate that is most likely to produce a single bolide observation in our observation time period. If bolides occur independently of each other, their occurrence can be described by a Poisson distribution. In this case, the probability of a single event occurring in time $T$ is given by $(rT)e^{-rT}$, where $r$ is the occurrence rate. This probability is a maximum when $r=1/T$. In this case, $T$ is 1300 seconds, so $r$ is 1 per 1300 seconds, or $\sim$24,000 per year.

Our observation was made from a distance of 80,000 km  and we estimate that the bolide we observed was due to an impactor of mass 250--5000 kg. While we would certainly be able to detect smaller impactors when the spacecraft is even closer to the planet, much of the observing time is when the spacecraft is further away. It is therefore difficult to associate the impact flux with a specific mass threshold; in the following discussion, we use both the 250 kg and 5000 kg values. 

\noindent\textbf{Text S2: Comparison of impact rate with previous studies}

\citeA{bland06} compiled a range of observational studies in order to produce a graph showing the rate of impact of meteoroids in the Earth's upper atmosphere as a function of meteoroid size.  There is a log-log relationship between impact rate and size. For meteoroids in the mass range 3--$10^{10}$ kg, the relationship is
\begin{equation}
    \log N = -0.926 \log m + 4.739
\label{eq:mass_rate}
\end{equation}
where $N$ is the number of impacts per year of mass $m$ or greater. Using this equation, the Earth undergoes $\sim$330 impacts of size \textgreater250 kg and $\sim$20 impacts of size \textgreater5000 kg each year. However, given the difference in orbit, mass and area, we clearly expect flux rates to be different on Jupiter. There have been two previous estimates of small impactor flux rates on Jupiter, one based on even smaller impactors \cite{cook81}, and one based on slightly larger impactors \cite{hueso18}. If we assume that the distribution of meteoroid sizes is the same for Jupiter as it is for the Earth, we can use Equation~\ref{eq:mass_rate} to compare our flux rate with these previous studies.

\citeA{cook81} used the camera on Voyager 1 to search for fireballs and detected one such event. Based on this, they estimated an impact rate of $1.6\times10^{-10}$ s\textsuperscript{-1}km\textsuperscript{-2} ($3\times10^{8}$ impacts per year) for impactors of mass \textgreater2.8 kg. Using Equation~\ref{eq:mass_rate}, we would expect \textgreater250 kg impactors to be 64 times less frequent than \textgreater2.8 kg impactors, i.e. the results of \citeA{cook81} suggest a rate of $5\times10^{6}$ impacts per year for impactors \textgreater 250 kg. Similarly, we would expect \textgreater5000 kg impactors to be 1000 times less frequent than \textgreater2.8 kg impactors, i.e. the results of \citeA{cook81} suggest a rate of $3\times10^{5}$ impacts per year for impactors \textgreater 5000 kg. Our estimated flux rate of 24,000 impacts per year is approximately 10--200 times lower than these values.

\citeA{hueso18} combine the results of several amateur observations over multiple years to estimate a impact rate for Jupiter for impactors in the $1\times10^{5}-1\times10^{6}$ kg range. Amateur observers have observed five fireballs in Jupiter's atmosphere in 2010--2017, and by estimating the effective time during which the amateur observing community takes measurements, \citeA{hueso18} estimate an impact rate of 10--65 events per year. Following the mass distribution given in Equation~\ref{eq:mass_rate}, this corresponds to a rate of 2,600--17,000 impacts per year for masses \textgreater 250 kg and a rate of 160--1,000 impacts per year for masses \textgreater 5000 kg. Our estimated flux rate of 24,000 impacts per year is higher than both of these values, but our lower mass estimate of 250 kg produces an impact rate that is significantly more consistent with \citeA{hueso18}. 

The previous two paragraphs compare three impact rate estimates: $3\times10^{8}$ per year for impactors \textgreater2.8 kg~\cite{cook81}, 24,000 per year for impactors \textgreater250--5000 kg (this paper) and 10--65 per year for impactors \textgreater$1\times10^{5}$ kg~\cite{hueso18}. If the mass distribution of impactors follows Equation~\ref{eq:mass_rate}, then our measured rate is lower than the value expected from \citeA{cook81} and higher than the value expected from \citeA{hueso18}. However, we note that all three rates have a significant amount of uncertainty as they are based on a very small number of impact observations: one each in the case of this paper and~\citeA{cook81} and five in the case of \citeA{hueso18}. While \citeA{hueso18} has a larger number of impacts, the effective duration of the observations is uncertain.


%
%

\end{document}